\documentstyle[12pt]{article}
\newfont{\larom}{cmbx10 scaled\magstep3}
\newfont{\bsan}{cmssbx10}

\begin{document}

\begin{center}
  {\larom On Modelling a Relativistic Hierarchical (Fractal) Cosmology by 
   Tolman's Spacetime -- I. Theory}  

  \vspace{10mm}
  {\Large Marcelo B. Ribeiro \\}
  \vspace{5mm}
  {\normalsize Astronomy Unit \\ School of Mathematical
  Sciences \\ Queen Mary and Westfield College \\ Mile End Road \\
  London E1 4NS \\ England}


  \vspace{10mm}
  {\bf ABSTRACT}
\end{center}
  \begin{quotation}
    \small

    In this work we examine a relativistic model for the observed 
    inhomogeneities of the Large Scale Structure where we make the hypothesis
    that this structure can be described as being a self--similar fractal
    system. The old Charlier concept of hierarchical clustering is
    identified with a fractal distribution and the problems raised by the use
    of fractal ideas in a relativistic model are discussed, as well as their
    relations to the Copernican and Cosmological Principles.
    Voids, clusters and superclusters
    of galaxies are assumed to be part of a smoothed--out fractal structure
    described by a Tolman solution. The basic concepts of the Newtonian model
    presented by Pietronero (1987) are reinterpreted and applied to this
    inhomogeneous curved spacetime. This fractal system is also assumed 
    to have a crossover to homogeneity which leads to a ``Swiss cheese''
    type model, composed by an interior Tolman metric and an exterior
    dust Friedmann solution. The Darmois junction conditions between the two
    spacetimes are calculated and we also obtain for the interior region
    the observational relations necessary to compare the model with
    observations. The differential equations of the interior spacetime
    are set up and we devise a numerical strategy for finding particular
    Tolman solutions representing a fractal behaviour along the past light
    cone.

    \vspace{3mm}

    \underline{Subject headings}: cosmology: theory -- galaxies:
    clustering -- large-scale structure of the universe -- relativity

  \end{quotation}
\vspace{6mm}
\newpage

\section{Introduction}
  \ \ \ The observational analysis of the CfA redshift survey made by de
  Lapparent, Geller and Huchra (1986) was the first clear confirmation that the
  Large Scale Structure of the Universe does not show itself as a smooth
  and homogeneous distribution of luminous matter as was thought earlier.
  Rather the opposite, since up to the limits of the observations presented
  in their article, the 3--D cone maps show a very inhomogeneous picture,
  with galaxies mainly grouped in clusters or groups alongside regions
  devoid of galaxies, virtually empty spaces with scales of the same order of
  magnitude as their neighbour clusters. More recent surveys, much deeper
  than the previous ones,  came to confirm those earlier findings presenting
  the Large Scale Structure as a complex mixture of interconnected voids,
  clusters and superclusters, observations that even led to the virtual
  discarding of some models which were based on the assumption that at
  the scale of these surveys, the Large Scale Structure would turn 
  into a homogeneous one (Saunders et al. 1991).

  \ \ \ If to see is to believe, the orthodox homogeneous picture seems to be
  in trouble when confronted with these observations, specially because
  a pattern appears to be common in all surveys: the deeper the
  probing is made, the more similar structures are observed and mapped,
  with clusters turning into superclusters and even bigger voids being
  identified.

  \ \ \ With respect to this pattern, two ideas seem to fit in. The first
  is the old concept of `hierarchical clustering' first advocated
  in the astronomical context by Charlier (1908, 1922) which states that
  galaxies join together to form clusters that form  superclusters which
  themselves are elements of super--superclusters and so on, possibly
  ad infinitum. The second and more recent concept is of `fractals',
  of which, for the present purpose, a rather loose tentative definition
  proposed by B. B. Mandelbrot seems to be adequate: ``A fractal is a
  shape made of parts similar to the whole in some way'' (see Feder 1988,
  p. 11).

  \ \ \ Hierarchical cosmology has been investigated in Newtonian 
  (Wertz 1971; Peebles 1974) and relativistic frameworks (Wesson 1978a,
  1979), but the first single fractal model 
  advanced as a description of the Large Scale Structure is due to
  Pietronero (1987, hereafter referred to only as Pietronero). Calzetti,
  Giavalisco and Ruffini (1988) followed similar arguments 
  and investigated further implications of the fractal hypothesis for
  galactic clustering statistics. There have also been attempts
  to measure the fractal dimension of the distribution of galaxies either
  by assuming a single fractal approach or a multifractal one (Balian and
  Schaeffer 1988; Deng, Wen and Liu 1988; Jones et al. 1988;
  Mart\'{\i}nez et al. 1990).

  \ \ \ This work is an attempt to generalize Pietronero's fractal model
  into a relativistic framework. It differs from Ruffini, Song and Stoeger 
  (1988) in that here we do not make use of a perturbation scheme. It is
  an exploratory model where some strong simplifying assumptions are
  made in order to
  avoid introducing unnecessary complications at this stage. In doing so,
  we shall assume that the large scale galactic clustering can
  be reasonably approximated by a single fractal and, hence, multifractals
  will not be treated here. We shall also assume relativistic dust
  solutions. This assumption enables us to model the smoothed--out fractal
  system through the general inhomogeneous dust solution due to
  R. C. Tolman (1934). We shall also consider a dust Friedmann
  spacetime as a background, as explained in section 2.

  \ \ \ In the next section is presented a very brief summary
  of Pietronero's main results needed in this work, as well as the
  identification and discussion of basic difficulties arising when trying
  to apply fractal ideas in General Relativity, including their relation
  with the Copernican and Cosmological Principles. We also
  discuss how we can get around these difficulties and build up a
  simple model. In section 3 the observational relations
  of a fractal model  in Tolman's spacetime are obtained and in section
  4 the junction conditions between Tolman and Friedmann spacetimes are
  discussed.  In section 5 the whole strategy and problems for solving
  numerically the differential equations of the model are exposed.
  The paper finishes with a concluding section.

\section{Hierarchical clustering and fractals}
  \ \ \ As mentioned in the previous section, Pietronero presented
  a model for the large scale distribution of galaxies where this
  distribution is assumed to form a self--similar
  fractal structure.  In this context, self--similarity means
  that a fractal consists of a system in which more and more structure
  appears at smaller and smaller scales and the structure at small scales is
  similar to the one at large scales (Mandelbrot 1983). It is,
  therefore, evident that fractals are simply a more precise version
  of the `scaling' idea behind Charlier's concept of hierarchical
  clustering. Earlier attempts 
  to model hierarchy started only with Charlier's hypothesis and, maybe,
  that is why all those models suffered a basic weakness: the lack of a
  precise mathematical definition for hierarchy. It is this difficulty
  that fractals are, it seems, able to successfully address. Basically
  fractals give a meaning to hierarchy.

  \ \ \ Further to the fractal hypothesis, Pietronero defines what he calls
  a `generalized mass--length relation' by starting from a point occupied
  by an object and counting how many objects are present within a volume 
  characterized by a certain length scale. For a deterministic self--similar
  distribution, we have that within a certain radius $d_0$, there are $N_0$
  objects; then within $d_1 = kd_0$ there are $N_1 = \tilde{k} N_0$ objects;
  in general, within $d_n = k^n d_0$ we have $N_n = \tilde{k}^n N_0$. 
  Generalizing this idea to a smooth relation, he then defines a relation
  between $N$ and $d$ of the type $N(d) = \sigma d^D$ where the fractal
  dimension $D = \log \tilde{k}/\log k$ depends only on the rescaling factors
  $k$ and $\tilde{k}$ and the prefactor $\sigma$ is related to the lower
  cutoffs $N_0$ and $d_0$, $\sigma = N_0/{(d_0)}^D$.

  \ \ \ Although fractals are essentially simple, their use in a
  relativistic framework is not so straightforward. The difficulties
  start with the recognition that Pietronero's relation $N \propto d^D$
  basically divides the spatial points of the system into two distinct
  categories: the points that belong to the fractal system where
  $N \propto d^D$  is valid and the ones that do not. In this sense
  each belonging point of the fractal system describes its remaining
  part by means of Pietronero's relation. In particular, any two geometrically
  identical portions of the fractal system carry identical number counts. A
  system with this property is called a `homogeneous fractal' (Mandelbrot 1983,
  p. 87), though the resulting distribution over the whole space is grossly
  inhomogeneous. The first difficulty can be understood if we remember that the
  Cosmological Principle states that all observers are indistinguishable. In
  other words, the Cosmological Principle is realised by a continuous
  group of symmetry imposed upon the points of our Riemannian manifold
  (see e.g. MacCallum 1983). Therefore, the fractal property of
  dividing the space into two different categories of points runs
  against the realisation of a continuous group of  symmetry on all
  points of the manifold and, consequently, a clash between fractals and the
  Cosmological Principle is all but unavoidable.

  \ \ \ Such a situation, therefore, leads us to a choice between two
  possibilities: if one wishes to keep the Cosmological Principle one is
  forced to give up fractals in cosmology. On the other hand, if one is
  willing to accept the empirical evidence and use fractals in cosmology
  one must adopt a weaker interpretation of the Copernican Principle
  (of no preferred points in the universe) which would be compatible
  and applicable to fractals. In this respect, Mandelbrot (1983, p. 205)
  advanced the `Conditional Cosmological Principle' which does not
  refer to all observers, but only to the material ones. That
  naturally leads to the hypothesis of a homogeneous fractal to describe
  galactic clustering possessing some symmetry
  around isolated material points which would form the fractal
  system. This hypothesis actually means dropping a continuous group
  of symmetry on all points of the manifold. By isolated points we mean
  points which have a neighbourhood  not containing other points of the
  same category.  Under this definition isolated points would form a subset
  of the Riemannian manifold. If the universe were finite, there would be a
  finite number of isolated points and this will in any case be true of
  the observable universe. In an infinite universe this number could be
  infinite.

  \ \ \ We can go a step further and assume a fractal possessing spherical
  symmetry around isolated points. Nevertheless, here again another difficulty
  comes out in that such a requirement would demand an overall property
  nonexistent in known geometries. Inhomogeneous spherically symmetric
  spaces certainly have one centre of symmetry and might also have two,
  which means that we would be giving up not only the Cosmological
  Principle but also the Copernican Principle in a cosmology with
  such geometries. Friedmann spacetime has spherical symmetry around
  infinite points and, hence, does not allow isolated points. Besides,
  its symmetries are such that Pietronero's relation cannot hold in its
  full generality, in principle $0~<~D~\leq~3$, and, thus, we can only
  conclude that a `Relativistic Fractal Cosmology' cannot be built
  within the Standard Friedmannian Cosmology. 
  
  \ \ \ The only way fractals could be seen within the context of
  the Standard Cosmology is if we remember that the Cosmological Principle
  has a statistical significance. That means the Cosmological Principle is, in
  practice, a statement that metric perturbations are small and this can
  be satisfied with density fluctuations $\delta \rho / \rho $ of some
  fractal types. Although this point of view may have an appeal to those
  who stick to Friedmannian Cosmology, it actually relegates fractals to
  nothing more than one possible type of local perturbations, a view
  already challenged by observations from the IRAS survey (Saunders
  et al. 1991).

  \ \ \ Departures from the Cosmological Principle are not new. Wesson
  (1978b) advanced one which is somewhat related to the discussion above
  in the sense that he sought a formulation of the Cosmological Principle 
  suitable to models where the density, pressure, etc, appear only in
  dimensionless functions solely dependent on the epoch.
  
  \ \ \ In addition to these geometrical difficulties, one could argue that
  the observations do not contradict the possible existence of an
  upper cutoff of the fractal system, beyond which the distribution
  becomes homogeneous, though Coleman, Pietronero and Sanders (1988) claim
  that there is no evidence for this cutoff in the CfA survey if a
  different from usual statistical analysis is carried out on it.

  \ \ \ Despite these difficulties and constraints, it is still possible
  to build up a simple relativistic fractal model if one adopts some sort of
  Einstein--Straus geometry (Einstein and Straus 1945, 1946), with the 
  interior solution consisting of the inhomogeneous Tolman spacetime and
  the exterior one of the dust filled Friedmann solution. In
  this way, the arbitrary functions of Tolman's solution can be used to 
  simulate a fractal system. 
  
  \ \ \ Modelling the Large Scale Structure in the form as described above 
  is a different way and new combination of looking at old ideas. ``Swiss
  cheese'' type models have proved to be popular in the examination of
  cosmological inhomogeneities
  (Lake 1980; Bonnor 1987 and references therein). Nevertheless, as far as
  we know the matching between Tolman and Friedmann using Darmois
  junction conditions was only briefly mentioned by Kantowski (1969),
  but without showing the calculations.
  The idea of using the arbitrariness in Tolman for simulation
  was already present in Bonnor (1972) in a more
  restricted model, though he did not solve the geodesic equation and,
  therefore, his simulation was over our present time hypersurface. As 
  will be shown next, in this work we shall develop the model along
  the past light cone where the observations are actually made and using
  Tolman's solution in its full generality, without restrictions. Relativistic
  hierarchical cosmological models were attempted by Wesson (1978a, 1979),
  but without the fractal concept his hierarchy became ill--defined.
  In addition, Bonnor did not fully express his model in
  terms of observational quantities, relating its density to the
  unobservable radius coordinate (something also done by Wesson)
  at constant time, wherein here we adopt
  the opposite approach. It is, however, the analytical complexity of Tolman's
  solution that actually prevented its development along these lines,
  demanding a numerical approach as will become clear in what follows.

\section{Tolman's solution as a fractal model for the distribution of 
  galaxies}
  \ \ \ We shall approach a relativistic generalization of Pietronero's model
  by assuming that Tolman's solution can be used as an approximation to 
  describe a fractal distribution of galaxies. Tolman (1934) obtained the
  general solutions of Einstein's equations for spherically symmetric dust in
  comoving coordinates which, in Bonnor's notation (Bonnor 1972), may be
  written (with $\Lambda = 0$ and $c = G = 1$)
    \begin{eqnarray}
      dS^2=dt^2-\frac{R'^2}{f^2}dr^2-R^2d\Omega^2, \ \  r\geq 0, \  R > 0
      \label{31}
    \end{eqnarray}
  where
    \begin{eqnarray} 
      d\Omega^2 \equiv d\theta^2 + \sin^2 \theta d\phi^2
      \label{32}
     \end{eqnarray} 
  is the usual metric on the unit sphere, $f$ is an arbitrary function of $r$
  only assumed to be of class $C^2$, i.e., having continuous second
  derivative, $R(r,t)$ satisfies
    \begin{eqnarray}  
      2R\dot{R}^2 + 2R(1-f^2) = F
      \label{33}
    \end{eqnarray}  
  and the proper density is given by
    \begin{eqnarray} 
      8 \pi \rho = \frac{F'}{2R'R^2}.
      \label{34}
    \end{eqnarray}   
  The dot means $\partial/\partial t$ and the prime means $\partial/\partial
  r$, and $F$ is an arbitrary function of $r$ also of class $C^2$.

  \ \ \ The solution of equation (\ref{33}) is known in the literature
  (Bonnor 1956,
  1974) and it has three distinct cases according as $f^2=1, \ f^2>1$ and
  $f^2<1$, these cases being termed respectively parabolic, hyperbolic and
  elliptic models (Bonnor 1974).

  \ \ \ In the parabolic models ($f^2=1$) the solution of equation (\ref{33}) is
    \begin{eqnarray} 
      R = \frac{1}{2} {(9F)}^{1/3} {(t+\beta)}^{2/3},
      \label{35}
    \end{eqnarray}
  where $\beta(r)$ being an arbitrary function assumed of class $C^2$.
  We shall need in further calculations a second partial derivative
  of equation (\ref{35}) and the first ones, which were obtained as follows:
    \begin{eqnarray}
      \dot{R} =  {\left[ \frac{F}{3(t+\beta)} \right] }^{1/3} ;
      \label{36}
    \end{eqnarray}
    \begin{eqnarray}
      R'=\frac{1}{3}  {\left( \frac{9F}{t+\beta} \right) }^{1/3} \left[
	 \frac{(t+\beta)}{2F} F' + \beta' \right] ;
      \label{37}
    \end{eqnarray}
    \begin{eqnarray}
      \dot{R}' = \frac{1}{9}  {\left( \frac{9F}{t+\beta} \right) }^{1/3} 
		   \left( \frac{F'}{F}-\frac{\beta'}{(t+\beta)} \right).
      \label{38}
    \end{eqnarray}

  \ \ \ In the hyperbolic models ($f^2>1$) the solution of equation 
  (\ref{33}) may be
  written in terms of a parameter $\Theta$,
    \begin{eqnarray} 
      R=\frac{F(\cosh 2\Theta-1)}{4(f^2-1)},
      \label{39}
    \end{eqnarray}
    \begin{eqnarray} 
      t+\beta=\frac{F(\sinh 2\Theta-2\Theta)}{4{(f^2-1)}^{3/2}}
      \label{310}
    \end{eqnarray}
  and these quantities' derivatives can be found as 
    \begin{eqnarray} 
      \dot{R}=\left( \frac{\sinh 2\Theta}{\cosh 2\Theta-1} \right) 
              \sqrt{f^2-1},
      \label{311}
    \end{eqnarray}

    \vspace{5mm}
    \[
      R'= \left[ \frac{1}{4(A-1){(f^2-1)}^2} \right] 
          \left[ (4A+B^2-6B\Theta-4)Fff' -
          \right.
    \]
    \begin{equation}
      \left. -2F'(A-B\Theta-1)(f^2-1)+4B\beta'{(f^2-1)}^{3/2} \right],
      \label{312}
    \end{equation}

    \vspace{5mm}

    \[ 
      \dot{R}'= \frac{1}{F(f^2-1) \left[ (3B^2+4)-A(B^2+4) \right]}
                \left\{ \sqrt{f^2-1} \left[ (5B-6\Theta)A -
		\right.
		\right.
    \]
    \[
      \left.    -B^3-5B+6\Theta \right] Fff'- \left[ F' \sqrt{f^2-1}
	        (B-2\Theta) -  \right.
    \]
    \begin{equation}
      \left. \left. -4 \beta' {(f^2-1)}^2 \right] (A-1)(f^2-1) \right\},
      \label{313}
    \end{equation} 
  where
    \begin{eqnarray} 
      A \equiv \cosh 2\Theta , \ \ \ \  B \equiv \sinh 2\Theta.
      \label{314}
    \end{eqnarray} 

  \ \ \ Finally, in the elliptic models $(f^2 < 1)$ a parameter $\Theta$ is
  again needed to write the solution of equation (\ref{33})
  \begin{eqnarray}
    R=\frac{F(1-\cos 2\Theta)}{4 {\mid f^2-1 \mid}},
    \label{315}
  \end{eqnarray}
  \begin{eqnarray}
    t+\beta= \frac{F(2\Theta-\sin 2\Theta)}{4{\mid f^2-1 \mid}^{3/2}},
    \label{316}
  \end{eqnarray}
  whose derivatives are
  \begin{eqnarray}
    \dot{R}= \left( \frac{\sin 2\Theta}{1-\cos 2\Theta} \right) \sqrt{\mid f^2
	      -1 \mid},
    \label{317}
  \end{eqnarray}

  \vspace{5mm}
  \[
    R' = \left[ \frac{1}{4(A-1){\mid f^2-1\mid}^2} \right] 
	 \left[ (4A-B^2+6B\Theta-4)Fff' +
	 \right.
  \]
  \begin{equation}
         \left.
	 +2 F' (A+B\Theta-1) \mid f^2-1\mid  -4B \beta' {\mid f^2-1 \mid}^{3/2}
	 \right],
    \label{318}
  \end{equation}

  \vspace{5mm}
    \[ 
      \dot{R}'= \frac{1}{F\mid f^2-1\mid \left[ A(B^2-4)+4-3B^2 \right]}
                \left\{ \sqrt{\mid f^2-1\mid} \left[ (5B-6\Theta)A +
		\right.
		\right.
    \]
    \[
      \left.    +B^3-5B+6\Theta \right] Fff'+ \left[ F' \sqrt{\mid f^2-1\mid}
	        (B-2\Theta) +  \right.
    \]
    \begin{equation}
      \left. \left. +4 \beta' {\mid f^2-1\mid}^2 \right] (A-1)\mid f^2-1\mid 
      \right\},
      \label{319}
    \end{equation} 
  where
    \begin{equation}
      A \equiv \cos 2\Theta, \ \ \ \ B \equiv \sin 2\Theta.
      \label{320}
    \end{equation}

  \ \ \ In order to make use of Tolman's models as descriptors of observations,
  it is necessary first of all to adopt the appropriate definition of distance
  of a radiating source, which in this case will be assumed to be the 
  `luminosity distance' as that is the definition generally used by
  observers in their data analysis. Its expression can be obtained by
  calculating first the `observer area distance' $r_0$ (see Ellis 1971;
  this is the same as the `corrected luminosity distance' of Kristian and
  Sachs 1966 and also the same as the `angular diameter distance' of Weinberg
  1972)
    \begin{equation}
      {(r_0)}^2=\frac{dS_0}{d\Omega_0}=\frac{R^2 d\theta \sin \theta d\phi}
	    {d\theta \sin d\phi}=R^2
      \label{321}
    \end{equation}
  in the spacetime (\ref{31}). Here $d\Omega_0$ is the solid angle subtended
  by a bundle of null geodesics diverging from the observer and $dS_0$ the 
  cross--sectional area of this bundle at some point. Further, it was shown
  by Ellis (1971) that the luminosity distance $d_l$ and the observer area
  distance are related by
    \begin{equation}
      {(d_l)}^2={(r_0)}^2{(1+z)}^4
      \label{322}
    \end{equation}
  which implies
    \begin{equation}
      d_l=R{(1+z)}^2
      \label{323}
    \end{equation} 
  in Tolman's spacetime. Here $z$ is the `redshift' of a source as measured
  by the observer.

  \ \ \ The next step in applying Pietronero's procedure in Tolman's spacetime
  is to obtain the expression for `number counts'. In any cosmological
  model if
  we consider a small affine parameter displacement $d\lambda$ at some point P
  on a bundle of past null geodesics subtending a solid angle $d\Omega_0$, and
  if $n$ is the number density of radiating sources per unit proper volume, then
  the number of sources in this section of the bundle is (Ellis 1971)
    \begin{equation} 
      dN={(r_0)}^2 d\Omega_0 \left[n(-k^a u_a) \right]_P d\lambda
      \label{324}
    \end{equation}
  where $k^a$ is the propagation vector of the radiation flux and $u^a$ is the
  4--velocity of the observer. Assuming a comoving observer $u^a = (1,0,0,0)$
  and that the past null geodesic is a radial one, given by
    \begin{equation}
      \frac{dt}{d\lambda}= - \left(\frac{R'}{f} \right) \frac{dr}{d\lambda},
      \label{325}
    \end{equation} 
  and also remembering spherical symmetry, equation (\ref{324}) becomes
    \begin{equation} 
      dN=4 \pi n \frac{R'R^2}{f} dr.
      \label{326}
    \end{equation} 
  We shall also assume that the sources are mostly galaxies, with rest masses
  of $M_G~\sim~10^{11} M_\odot$ and, therefore, equation (\ref{34}) allows us to
  write
    \begin{equation}
      n = \frac{\rho}{M_G} = \frac{F'}{16 \pi M_G R' R^2}.
      \label{327}
    \end{equation}  
  
  \ \ \ Once we substitute equation (\ref{327}) into equation (\ref{326})
  and integrate the
  latter, we obtain the number $N_c(r)$ of sources which lie at radial
  coordinate distances less than $r$ as seen by the observer at $r=0$
    \begin{equation}
      N_c(r) = \frac{1}{4M_G} \int_C \frac{F'}{f} dr,
      \label{328}
    \end{equation} 
  where the integration is made along the curve $C$ formed
  by the past light cone parametrized by $r$. Two notes should be made
  about the equation above. Firstly, the affine parameter $\lambda$
  becomes implicit, a fact which brings advantages in carrying out
  numerical calculations. Secondly, if we let $t(r)$ be the solution of
  the geodesic $C$, which is given by equation (\ref{325}), we can see 
  that although equation (\ref{328}) does not have the time coordinate
  explicitly in the right hand side, the integration is along  the
  geodesic where $R = R[r,t(r)]$. That is because equation (\ref{325})
  was used in the derivation of equation (\ref{328}). 

  \ \ \ Now in order to make the appropriate definition of density applicable
  to a fractal model, we will follow Wertz (1971) and Bonnor (1972) and
  distinguish between a `volume density' $\rho_v$ obtained by averaging
  over a sphere of given volume and the `local density' $\rho$ given by
  equation (\ref{34}). Nevertheless, our definition of volume density is
  different
  from the latter inasmuch as in this model we use the luminosity 
  distance as our definition of distance, a fact that basically means that we
  observe distances in a curved spacetime as if this spacetime were a
  Euclidean one. In other
  words, $d_l$ is the distance which the source would be at if it were
  stationary in a Euclidean space. In this sense, therefore, the volume of
  the sphere which contains the sources may be written as
    \begin{equation}
      V(r)=\frac{4}{3} \pi {(d_l)}^3 = \frac{4}{3} \pi R^3 {(1+z)}^6
      \label{329}
    \end{equation} 
  and the volume density is given by
    \begin{equation}
      \rho_v(r)=\frac{M_GN_c(r)}{V(r)}=\frac{3}{16\pi R^3 {(1+z)}^6 }
		\int_C \frac{F'}{f}dr.
      \label{330}
    \end{equation}  

  \ \ \ This expression  merely states the volume density in 
  Tolman's spacetime and does not contain by itself any relationship to a
  fractal distribution of dust. Therefore, following Pietronero's hypothesis
  for a self--similar fractal distribution\footnote{Here the self--similarity
  due to fractals should not be confused with the one due to homothetic
  Killing vectors. The latter is discussed in Cahill and Taub (1971).},
  if within a
  certain radius $(d_l)_0$ there are  $(N_c)_0$ objects and then within
  $(d_l)_1$ there are $(N_c)_1$ objects, we can then write a
  smoothed--out relation between $N_c$ and $d_l$ as 
    \begin{equation} 
      N_c=\sigma {(d_l)}^D,
      \label{331}
    \end{equation}  
  where $\sigma$ is a constant related to the lower cutoffs $(N_c)_0$ and
  $(d_l)_0$ of the distribution and $D$ is its fractal dimension that can
  be noninteger. This is the natural generalization of Pietronero's
  definition originally made in a Newtonian context.

  \ \ \ We must point out that the adoption of equation (\ref{331}) is
  the obvious
  thing to do if one wishes to follow the astronomical procedure and
  compare the model with observations. Nevertheless, fractal dimensions
  have so far been defined in Euclidean spaces and it is not at all
  clear whether equation (\ref{331}) is the most appropriate definition to 
  take in curved spacetimes. We can see a possible shortcoming if we
  remember that it is usually assumed that in a homogeneous distribution
  $D \cong 3$ (Mandelbrot 1983, Pietronero) and one could argue that
  this would be the value to be found for Friedmann.  However, Friedmann
  spacetime is homogeneous at constant time coordinates and when we
  integrate along the past light cone, going through hypersurfaces of
  $t$ constant with each one having different values for the density, it
  should not be so surprising if $D$ departs from the value 3 even in a
  spatially homogeneous spacetime.

  \ \ \ From equation (\ref{331}) and also considering equations (\ref{329})
  and (\ref{330}),
  it is possible now to compute the volume density for a sphere of certain
  radius that contains a portion of the fractal distribution:
    \begin{equation} 
      \rho_v=\frac{3\sigma M_G}{4 \pi} {(d_l)}^{-\gamma}, \ \ \ \gamma=3-D.
      \label{332}
    \end{equation} 
  This is the same sort of expression as obtained by de Vaucouleurs (1970)
  when he argued in favour of a hierarchical cosmology. 
  If we now take the
  volume density (\ref{330}) and substitute into equation (\ref{332}) we get
    \begin{equation}
      \int_C \frac{F'}{f} dr=4 \sigma M_G {\left[ R {(1+z)}^2 \right] }^D.
      \label{333}
    \end{equation} 
  This is the condition that the three arbitrary functions $f(r), F(r)$ and
  $\beta(r)$ must satisfy such that a fractal distribution of galaxies is
  simulated in Tolman's spacetime. We can call equation (\ref{333}) the 
  `self--similar condition' as it is clearly the particular case of the general
  equation (\ref{331}) when applied to Tolman's solution. 

  \ \ \ As the final issue before the end of this section,  
  although the redshift is essential in all previous
  expressions, it has not been explicitly calculated for the spacetime under
  consideration. In order to do so, let us start with the general expression
  for the redshift (see e.g. Ellis 1971)
    \begin{equation} 
      1+z=\frac{{(u^a k_a)}_{\rm source}}{{(u^b k_b)}_{\rm observer}}.
      \label{334}
    \end{equation}
  We shall assume that both source and observer are comoving and, hence, 
  equation (\ref{334}) becomes
    \begin{equation}
      1+z= { \left. \frac{dt}{d\lambda}\right| }_{\lambda=\lambda_0} 
	   { \left( {\left. \frac{dt}{d\lambda}\right| }_{\lambda=0}
	   \right) }^{-1} 
      \label{335}
    \end{equation} 
  where $\lambda_0$ is any value taken by the affine parameter $\lambda$ along
  the geodesic.
  We shall make use of the condition that the spacetime should be regular at the
  spatial origin and, therefore, when $r \rightarrow 0, \  R=r, \  f=1,
  \  R'=1, \  F=0$ (Bonnor 1974). These conditions together with equation
  (\ref{325}) allow us to write
    \begin{equation}  
      1+z= { \left. \frac{R'}{f} \frac{dr}{d\lambda}\right| }
	   _{\lambda=\lambda_0} 
      	   { \left( { \left. \frac{dr}{d\lambda}\right| }_{\lambda=0}
	   \right) }^{-1}.         
      \label{336}
    \end{equation}
  \ \ \ Following an idea suggested by M. A. H. MacCallum, the right hand
  side of equation (\ref{336}) can be calculated by starting with the
  Lagrangian of the radial metric
    \begin{equation} 
      L={\left( \frac{dS}{d\lambda} \right) }^2 =
	{\left( \frac{dt}{d\lambda} \right) }^2 -
	{\left( \frac{R'}{f} \frac{dr}{d\lambda} \right) }^2.
      \label{337}
    \end{equation} 
  The Lagrange equations of second kind
    \[
      \frac{d}{d\lambda} \frac{\partial L}{\partial \dot{x}^{\nu}} -
      \frac{\partial L}{\partial x^\nu} = 0, \ \ \ 
      \dot{x}^{\nu} = \frac{dx^\nu}{d\lambda}
    \]
  applied to equation (\ref{337}) result in
    \begin{equation}
      \frac{d^2t}{d\lambda^2}+ {\left( \frac{dr}{d\lambda} \right)}^2
	    \frac{R' \dot{R'}}{f^2} = 0, 
      \label{338}
    \end{equation} 

    \vspace{5mm}
    \begin{equation} 
      \frac{d^2r}{d\lambda^2}+\frac{1}{R'}
	    {\left( \frac{dr}{d\lambda} \right)}^2 \left( R'' -
	    \frac{f'R'}{f} \right) + 2 \frac{dr}{d\lambda}
	    \frac{dt}{d\lambda} \frac{\dot{R'}}{R'} = 0 .
      \label{339}
    \end{equation}
  Here in the second equation the assumption that ${(R')}^2 \neq 0$ was made
  (otherwise $g_{rr} = 0$).\footnote {Actually the boundary surfaces on
  which $R' = 0$ are shell crossings, where the density $\rho$ diverges
  and the region beyond has negative density. They indicate a
  breakdown of the basic assumptions of the Tolman metric (see Hellaby
  and Lake 1985, 1986 for details).} Considering the radial null geodesic
  it is possible to integrate equations (\ref{338}) and (\ref{339}) once,
  obtaining
    \begin{equation} 
      \frac{dt}{d\lambda} = \frac{1}{I+ C_1},
      \label{340}
    \end{equation}  
    \vspace{4mm}
    \begin{equation}
      \frac{dr}{d\lambda} = {\left[ \int \left( \frac{R''}{R'} -
      \frac{f'}{f} - \frac{2 \dot{R'}}{f} \right) d\lambda + C_2 \right] }^{-1}
      \label{341}
    \end{equation}  
  where
    \begin{equation}
      I \equiv \int \frac{\dot{R'}}{R'} d\lambda
      \label{342}
    \end{equation}    
  and $C_1, \ C_2$ are two integration constants whose relationship can be found
  by substituting equations (\ref{340}) and (\ref{341}) back into the geodesic:
    \begin{equation}
      \int \left( \frac{R''}{R'} - \frac{f'}{f} - \frac{2 \dot{R'}}{f} 
      \right) d\lambda + C_2 = - \frac{R'}{f} (I+C_1).
      \label{343}
    \end{equation} 
  This equation is  valid for any $\lambda$, including the point $\lambda=0$
  ($\lambda$ is taken to be zero at $r=0$) where the regular conditions make
  equation (\ref{343}) become
    \begin{equation}
      C_2=-C_1.
      \label{344}
    \end{equation}   
  The same conditions substituted into equation (\ref{340}) lead to
    \begin{equation}  
      {\left. \frac{dt}{d\lambda} \right| }_{\lambda=0} = \frac {1}{C_1}.
      \label{345}
    \end{equation} 
  However, as our observations are along the past null geodesic, the natural
  choice for $C_1$ is
    \begin{equation}
      C_1=-1 \ \ \ \ \Longrightarrow  \ \ \ \ C_2=1,
      \label{346}
    \end{equation}   
  which considering equation (\ref{343}) implies that equations (\ref{340})
  and (\ref{341}) may be written as
    \begin{equation}
      \frac{dt}{d\lambda}= \frac{1}{I-1},
      \label{347}
    \end{equation}  

    \vspace{3mm}
    \begin{equation} 
      \frac{dr}{d\lambda}= \frac{f}{(1-I)R'}.
      \label{348}
    \end{equation} 

  \ \ \ The redshift can, therefore, be calculated once we again make use of
  the regularity conditions on equation (\ref{348}) to get
    \begin{equation}  
      {\left. \frac{dr}{d\lambda} \right| }_{\lambda=0} =1
      \label{349}
    \end{equation}  
  that substituted into equation (\ref{336}), together with equation 
  (\ref{348}), gives
  \vspace{2mm}
    \begin{equation}
      z = \frac {I}{1-I}.
      \label{350}
    \end{equation} 
  \vspace{2mm}
  \ \ \ The integral $I$ still explicitly contains the affine parameter,
  which can be made implicit by considering equation (\ref{348}) and
  differentiating equation (\ref{342})  
    \begin{equation}
      \frac{dI}{dr}= \frac{\dot{R'}}{f} (1-I).
      \label{351}
    \end{equation} 
  Hence, the redshift in equation (\ref{350}) needs the solution of the
  differential equation above for the values of $I$. 

  \ \ \ As a final remark, it is of great numerical advantage that 
  equation (\ref{351}) is written only in terms of the radial
  coordinate $r$ and, therefore, can be solved simultaneously with the past
  radial null geodesic (\ref{325}). In this form the redshift becomes an
  implicit function of $r$ only, $z = z[I(r)]$.

\section{The matching to a dust Friedmann exterior}
  \ \ \ As discussed in section 2, the fractal system is assumed to have
  a crossover to homogeneity, which will be represented in this model by
  the matching between the inhomogeneous Tolman metric and the homogeneous
  Friedmann one. In order to achieve a smooth transition it is necessary
  to solve the junction conditions for the two metrics. In this case this 
  is a straightforward calculation in view of the fact that both metrics are
  comoving dust filled spherically symmetric spacetimes.

  \ \ \ Let us start by writing the Friedmann metric as
    \begin{equation}
      d{\overline{S}} \ ^2=dT^2-a^2(T)\left[dx^2+g^2(x)
			d{\overline{\Omega}} \ ^2 \right]
      \label{m1}
    \end{equation}
  where
    \begin{eqnarray*}
      d{\overline{\Omega}} \ ^2 \equiv d{\overline{\theta}} \ ^2+\sin^2
      {\overline{\theta}} d{\overline{\phi}} \ ^2,
    \end{eqnarray*}
  \vspace{3mm}
  \[
    g(x) = \left\{ \begin{array}{ll}
		   \sin x,     &    K = +1, \\
		   x,          &    K = 0, \\
		   \sinh x,    &    K = -1, 
		   \end{array}
          \right.
  \]
  and $a(T)$ satisfies the Friedmann equation
    \begin{equation}
      \dot{a}^2 = \frac{8 \pi}{3} \mu a^2 - K.
      \label{m2}
   \end{equation}
  Here the dot means $\partial/\partial T$ and the prime $\partial/
  \partial x$, and $\mu$ is the dust density.

  \ \ \ Let $\Sigma$ be a hypersurface which separates Riemannian
  spacetime into two four--dimensional manifolds $V^-$ and $V^+$
  (Israel 1966, 1967). Here $V^-$ is the interior Tolman metric and
  $V^+$ the exterior Friedmann one. The hypersurface $\Sigma$ is then
  defined by
    \begin{equation}
      \Sigma_- = r - \Sigma_0 = 0, \ \ \ \ \ \ \Sigma_+ = x - \Sigma_0 = 0
      \label{m3}
    \end{equation}
  where the indexes + and -- mean the approach to $\Sigma$ is from
  $V^-$ or $V^+$, and $\Sigma_0$ is the constant which defines the end
  of the Tolman cavity.

  \ \ \ The Darmois junction conditions state that $V^-$ and $V^+$
  match across $\Sigma$ if the first and second fundamental forms of
  $\Sigma$ are identical (Bonnor and Vickers 1981). As $V^+$ and $V^-$
  are  spherically symmetric, it is natural to take the intrinsic
  metric to $\Sigma, \ {dS_\Sigma}^2 \equiv g_{\alpha \beta}
  d\xi^\alpha d\xi^\beta, \ (\alpha = 0,2,3)$, also to be spherically
  symmetric, where $\xi^\alpha$ are the intrinsic coordinates of
  $\Sigma$. In this case $\xi^\alpha = x_+^\alpha = x_-^\alpha$ and,
  hence, $\xi^0 = t=T, \ \xi^2=\theta=\overline{\theta}, \ 
  \xi^3=\phi=\overline{\phi}$. Thus the first fundamental form
  identity of $\Sigma$, ${dS_-}^2 = {dS_+}^2$, leads to
    \begin{equation}
      R = ag \ \ \ \ \ \ \ \ \ {\rm on} \ \ \  \Sigma.
      \label{m4}
   \end{equation}
  \vspace{1mm}

  \ \ \ The unit normals to $\Sigma, \ {n_a}^{\pm} = {\left[ \Sigma_{,a}
  {\left( -g^{bc} \ \Sigma_{,b} \Sigma_{,c} \right)}^{-1/2}
  \right]}^{\pm},$ directed from $V^-$ to $V^+$, ($a$ = 0,1,2,3), are
  needed to calculate the condition for continuity of the second
  fundamental form or extrinsic curvature. For $V^-$ and $V^+$ they are
  respectively
    \begin{equation}
      {n_b}^- = \frac{R'}{f} \ \delta_b^1, \ \ \ \ \ \ {n_c}^+ = a \
      \delta_c^1.
    \end{equation}
  The extrinsic curvature $K_{ab} = n_{a;b}$ on $\Sigma$ takes the form
    \begin{equation}
      K_{\alpha \beta}= \frac{\partial x^a}{\partial \xi^\alpha}
      \frac{\partial x^b}{\partial \xi^\beta} K_{ab}
    \end{equation}
  and the explicit calculation of the condition ${K_{\alpha \beta}}^- =
  {K_{\alpha \beta}}^+$ gives
    \begin{equation}
      Rf = agg' \ \ \ \ \ \ \ \ \ {\rm on} \ \ \ \Sigma.
      \label{m5}
    \end{equation}
  Substituting the first condition into the second leads to
    \begin{equation}
      f = g' \ \ \ \ \ \ \ \ \ {\rm on} \ \ \ \Sigma.
      \label{m6}
    \end{equation}

  \ \ \ We can check whether these results are correct if we remember
  that both spacetimes are equivalent (spherically symmetric and dust
  filled) and, therefore, Tolman metric should change to the Friedmann one
  on $\Sigma$. This is easily verified if we substitute equation (\ref{m4}), its
  radial coordinate derivative and equation (\ref{m6}) into equation (\ref{31}).

  \ \ \ The junction conditions above have also an effect on the
  gravitational mass within the Tolman cavity. If we interprete
  equation (\ref{33}) as a total energy equation (Bondi 1947), we may define the
  gravitational mass inside $r$ as
    \begin{equation}
      m(r) \equiv \frac{F(r)}{4} = \int_0^r 4 \pi \rho R' R^2 dr
      \label{m7}
    \end{equation}
  which implies that equation (\ref{33}) may be written as
    \begin{equation}
      \frac{\dot{R}^2}{2} - \frac{m}{R} = \frac{1}{2} (f^2-1).
      \label{m8}
    \end{equation}
  Therefore, the total gravitational mass trapped by $\Sigma$ within the
  Tolman cavity is
    \begin{equation}
      M = m(\Sigma_0) = \int_0^{\Sigma_0} 4 \pi \rho R' R^2 dr.
      \label{m9}
    \end{equation}
  If we apply the junction conditions to the equation above we get
    \begin{equation}
      M=\int_0^{\Sigma_0} 4\pi \rho R'R^2dr=\int_0^{\Sigma_0}4\pi \mu a^3 g^2 g'
      dx = \frac{4 \pi}{3} \mu a^3 g^3(\Sigma_0) = \overline{M}
      \label{m10}
    \end{equation}
  where $\overline{M}$ is the gravitational mass of the Friedmann metric
  for the region $0 < x \leq \Sigma_0$. Therefore, the matching
  restricts the mass inside $V^-$. The gravitational mass
  must be the same as if the whole spacetime were Friedmannian and the
  Tolman cavity were never there. If there is any overdensity within the
  cavity, there must also be underdensity before the uniform region is
  reached, in order that the average densities will be the same.

  \ \ \ This constraint appears to imply that Einstein--Straus like
  geometries are too restricted to be used for understanding the nature
  of the real inhomogeneous universe, as the inhomogeneities will be
  severely restricted in a way that could be taken to be unnecessary.
  However, it has already been pointed out by Ellis and Jaklitsch (1989)
  that the matching can be used as a `fitting condition' specifying what
  is an appropriate Friedmann model to use as a background in a given
  lumpy universe model. In other words, if we can measure the mass
  distribution in our neighbourhood, that tells us whether the Friedmann
  background has enough mass to be a closed or open model. Hence, the
  matching conditions are interpreted not as a handicap but as advantageous
  cosmological fitting conditions, ensuring that the Friedmann model
  overall mass is correctly adapted to the inhomogeneous universe.

\section{Numerical methods}
  \ \ \ In section 3 the necessary expressions for modelling a
  fractal dust in Tolman's spacetime were developed and it was shown that
  excluding the number counting, all other relevant observational relations
  can be computed if we know the
  solutions of the two linear first order ordinary differential equations:
  the radial past null geodesic and the equation for the redshift
    \begin{equation}
      \frac{dt}{dr}=- \frac{R'}{f}, \ \ \ \  \ \ \ \ \frac{dI}{dr}=(1-I) 
      \frac{\dot{R'}}{f}.
      \label{41}
    \end{equation}  

  \ \ \ A brief inspection of the expressions for $R'$ and $\dot{R'}$
  shows  that
  an attempt to find an analytical solution for these equations is virtually
  hopeless, specially in the elliptic and hyperbolic models and, hence, a
  numerical approach is made necessary. Let us suppose that $t(r)$ is the
  solution of the geodesic and $I(r)$ of the equation for the redshift. The
  observations lie along the past light cone and in order to compare the
  numerical results with them, $R'$ and $\dot{R'}$ must be evaluated
  along the geodesic. Therefore, we must compute $R=R \left[ r,t(r) \right]$
  and its derivatives, which means that $I(r)$ can only be found if $t(r)$ is
  already known.

  \ \ \ We shall assume that `here and now' is defined by $r=0,
  \ t=0, \ 
  \lambda=0$, definition which implies the initial conditions for (\ref{41})
  as being
    \begin{equation}  
      t(0)=0, \; \ \ \ \ \ \ \ I(0)=0.
      \label{42}
    \end{equation}   
  This assumption, however, runs into trouble because due to the regular
  condition $F(0)=0$, at the origin of the elliptic and hyperbolic models the
  parameter $\Theta$ remains undefined. This difficulty can be overcome if we
  make the hypothesis that the metric remains flat from $r=0$ till some small
  value $r=\varepsilon$, and beyond it the spacetime changes to a curved one.
  Hence, we replace the initial conditions (\ref{42}) by
    \begin{equation} 
      t(\varepsilon)=- \ \varepsilon, \; \ \ \ \ \ \ \ I(\varepsilon)=0.
      \label{43}
    \end{equation}  

  \ \ \ In the previous sections it was explicitly assumed that the 
  fractal dust under consideration has a lower cutoff associated with
  the constant $\sigma$ of equation (\ref{331}), below which this structure is
  no longer observed. At the Galactic level no fractal distribution is
  observed and, therefore, we can naturally assume that this structure 
  starts at least at the scale of the Local Group, which would mean
  taking $\varepsilon \stackrel{<}{\sim}$ 1 Mpc.

  \ \ \ The goal of modelling Tolman's solution to a fractal distribution
  is to make use of the freedom of the arbitrary functions in order to find
  out particular functions $f(r), \  F(r), \  \beta(r)$ such
  that the volume density takes the de Vaucouleurs' density power law
  (\ref{332}). The self--similar condition 
  (\ref{333}) is of little practical use because its right hand side cannot be
  computed analytically. In these circumstances, the following numerical
  strategy was devised: we carry out the discretisation of the
  radial coordinate $r_i \ (i=1,2,\ldots,n; \: \varepsilon \leq r \leq 
  \Sigma_0 $) and for each set of points $r_i, \ t_i, \ I_i$ and $r_{i+1}$
  we calculate $t_{i+1}$ then $I_{i+1}$ through some
  numerical algorithm for solving ordinary
  differential equations. Also knowing $r_i$, equation (\ref{328}) permits the
  computation of $N_{c_i}$ by means of a numerical quadrature. In the
  elliptic and hyperbolic cases it is also 
  necessary to use a root--finding algorithm to evaluate $\Theta_i$. With
  these results it is possible to compute the observational quantities
  $d_{l_i}, \ \rho_{v_i}$ and $z_i$ through equations (\ref{323}), 
  (\ref{330}) and (\ref{350}) respectively. These values immediately allow us to
  plot graphics relevant to observations like number counting versus redshift.

  \ \ \ As these calculations will produce a great quantity of numbers, 
  it is necessary here a direct method of checking  whether a fractal
  distribution was modelled, specially because we might not easily see
  a true power law like expression for the volume density against the
  luminosity distance. For this purpose, we can simply take the logarithm
  of equation (\ref{332})
    \begin{equation}
      \log \rho_v = a_1 + a_2 \log d_l
      \label{44}
    \end{equation}
  and carry out a linear fitting over the points obtained through numerical
  integration.  Naturally, at each integration a particular set of functions
  $f(r), \ F(r), \ \beta(r)$, is chosen beforehand and if the fitting is
  successful (measured by an acceptable goodness of fit), that is, if the
  results have a linear form given by equation (\ref{44}) with negative
  slope, a fractal distribution of dust was modelled by a particular Tolman's
  spacetime. If the fitting is not successful a new attempt is made with a
  different set of functions. That method is tantamount to a numerical 
  simulation procedure for modelling a fractal distribution of dust by
  Tolman's spacetime.

  \ \ \ Once the fitting is successful, the fractal dimension $D$ and the
  constant $\sigma$ can be found directly from the fitted constants in 
  equation (\ref{44}) as
    \begin{equation}
      D  =  a_2+3 
      \label{45}  
    \end{equation}
  and             
    \begin{equation}
      \sigma   =  \frac{4 \pi}{3 M_G} \exp(a_1).
      \label{46}
    \end{equation}

  \ \ \ As stated above, in the hyperbolic and elliptic models for each
  $t_i(r_i)$, $\beta_i(r_i)$, $F_i(r_i)$, $f_i(r_i)$ we need to find 
  the root $\Theta_i$ of equations (\ref{310}) or (\ref{316}) in order to be
  able to evaluate $R_i$, $\dot{R}_i$, $R'_i$ and $\dot{R'}_i$. That is done
  numerically by finding an
  interval where the root lies, then using some algorithm to hunt it
  down. That interval obviously must be limited to the physical regions
  of the spacetime under consideration and, therefore, the following
  remark must be made. The function $\beta(r)$ determines the local time
  at which $R=0$ and, consequently, the hypersurface $t+\beta=0$ is a
  surface of singularity. In view of this the physical region to be
  considered is defined by $t+\beta > 0$.

  \ \ \ Bearing this point in mind, we can now proceed with the bracketing of
  the roots. In the elliptic case due to the boundness of the sine
  function it is easy to see that
    \begin{equation}
      \frac{4}{F} ( t+\beta ) {\mid f^2-1 \mid }^{3/2} -1 
      \leq  2 \Theta  \leq \frac{4}{F} ( t+\beta ) 
      {\mid f^2-1 \mid }^{3/2} +1.
      \label{f1}
    \end{equation}
 
  \ \ \ The hyperbolic case is a bit more complicated as  the hyperbolic
  sine is not bounded. Let us write equation (\ref{310}) as
    \begin{equation}
      G(\Theta) = \sinh 2 \Theta - 2 \Theta - \frac{4}{F}(t+ \beta)
      {(f^2-1)}^{3/2} = 0.
      \label{f2}
    \end{equation}
  The function $G(\Theta)$ changes sign within the interval $ \left[
  G(0), G(+ \infty) \right] $ which is where the root lies $(F \geq 0$
  otherwise we would have negative gravitational mass).
  As $G(0) < 0$ and $G(+ \infty) >0$, the change of sign
  occurs when the inequality $G(\Theta) >0$ is satisfied for $\Theta >
  0$. Using a power series expansion for $\sinh 2 \Theta$, this
  inequality can be written as
    \begin{equation} 
      \frac{{(2 \Theta)}^3}{3!} + \frac{{(2 \Theta)}^5}{5!} +
      \frac{{(2 \Theta)}^7}{7!} + \ldots > \frac{4}{F} (t+ \beta )
      {(f^2-1)}^{3/2}.
      \label{f3}
    \end{equation}   
  If $\Theta \geq 1$ the inequality will be satisfied provided the
  smallest term of the series is bigger than the right hand side of
  equation (\ref{f3}). If $0 < \Theta < 1$ the first term of the series will
  dominate and the inequality will be satisfied provided this first term
  is bigger than the right hand side of equation (\ref{f3}). In short, for the
  hyperbolic models the root of equation (\ref{310}) lies within the interval
    \begin{equation}
     0 < \Theta \leq {\left[ \frac{3}{F} (t+ \beta ){(f^2-1)}^{3/2}
     \right] }^{1/3}.
     \label{f4}
    \end{equation}   

\section{Conclusion}
  \ \ \ In this work we have proposed a relativistic hierarchical
  (fractal) cosmology in the inhomogeneous Tolman spacetime based on the
  reinterpretation and relativistic generalization of Pietronero's
  Newtonian model. We have assumed that the large scale distribution of
  galaxies forms a homogeneous fractal system and discussed how fractals
  give a new and precise meaning to Charlier's concept of
  hierarchical clustering. We concluded that the fractal property of
  dividing the space in points of different categories clashes with the
  Cosmological Principle, a fact which led us to seek a weaker interpretation of
  the Copernican Principle. In doing so we have assumed Mandelbrot's
  Conditional Cosmological Principle and made the hypothesis of a
  fractal with the property of being spherically symmetric around 
  isolated points. Such a fractal, however, demands an overall
  property nonexistent in known geometries.
  
  \ \ \ Considering these difficulties, we have advanced a simple
  exploratory model compatible with the Conditional Cosmological Principle
  by adopting a version of Einstein--Straus geometry consisting of
  an interior inhomogeneous Tolman spacetime and an exterior Friedmann
  one. Our fractal system is smoothed--out and has an upper cutoff
  which coincides with the
  end of the Tolman cavity. We have obtained the observational relations
  for the Tolman spacetime necessary to compare the model with the
  astronomical observations, namely the luminosity distance, number
  counts, volume density (average density) and redshift. We have also
  found a self--similar condition which the arbitrary functions of Tolman
  spacetime must satisfy in order to simulate a fractal dust. The Darmois
  junction conditions between the two spacetimes were also calculated.

  \ \ \ The differential equations necessary for evaluating the observational 
  relations were set up and we have discussed a numerical approach for
  solving these equations, inasmuch as that is virtually impossible
  to be done analytically. In this respect the numerical method consists of
  choosing particular Tolman's solutions and carrying out a linear
  fitting over the points obtained through numerical integration. That
  aims to see whether or not these particular solutions obey a de
  Vaucouleurs like power law
  relation for the volume density and then to find the fractal dimension
  of the distribution. Finally, in the physical region of the Tolman
  spacetime we found the interval where the parameter $\Theta$
  lies in the elliptic and hyperbolic cases.
  
  \ \ \ The numerical results of this model are the subject of a
  forthcoming paper.

  \vspace{15mm}
  \begin{flushleft}
  {\large \bf Acknowledgments}
  \end{flushleft}
  \vspace{5mm}

  \ \ \ First of all I would like to express my warm gratitude to 
  M. A. H. MacCallum for his guidance during this research and the
  numerous discussions and comments in all stages of this work, for
  calling my attention to fractals and for his suggestion credited in
  the body of this paper. In addition I am grateful to 
  W. B. Bonnor for discussions and comments and for calling my attention to 
  Einstein--Straus geometry. I also want to thank A. B. Burd for discussions,
  P. S. Wesson for a suggestion concerning his early papers, B. J. T. Jones
  whose criticisms and opposition prompted a clearer exposition of the
  underlying hypothesis of this paper and the referee for helpful
  comments. Finally, I am also grateful to the agency CAPES of Brazil for
  financial support.

\vspace{15mm}       
\begin{flushleft}
{\large \bf References}
\end{flushleft}
  \begin{description}
    \item  Balian, R., and Schaeffer, R. 1988, Ap. J. (Letters), 335, L43.
    \item  Bondi, H. 1947, M. N. R. A. S., 107, 410.
    \item  Bonnor, W. B. 1956, Zeit. f\"{u}r Astrophysik, 39, 143.
    \item  Bonnor, W. B. 1972, M. N. R. A. S., 159, 261.
    \item  Bonnor, W. B. 1974, M. N. R. A. S., 167, 55.
    \item  Bonnor, W. B. 1987, Ap. J., 316, 49.
    \item  Bonnor, W. B., and Vickers, P. A. 1981, Gen. Rel. Grav., 13, 29.
    \item  Cahill, M. E., and Taub, A. H. 1971, Commun. Math. Phys., 21, 1.
    \item  Calzetti, D., Giavalisco, M., and Ruffini, R. 1988, Astr. Ap.,
	   198, 1.
    \item  Charlier, C. V. L. 1908, Ark. Mat. Astron. Fys., 4, 1.
    \item  Charlier, C. V. L. 1922, Ark. Mat. Astron. Fys., 16, 1.
    \item  Coleman, P. H., Pietronero, L., and Sanders, R. H. 1988, Astr. Ap.,
	   200, L32.
    \item  Deng, Z.-G., Wen, Z., and Liu, Y.-Z. 1988, in Large Scale
           Structures of the Universe, Proc. of the 130th IAU Symposium, eds.
           Audouze, J., Pelletan, M.-C., and Szalay, A. (Dordrecht: Kluwer
           Academic Publishers), p. 555.
    \item  Einstein, A., and Straus, E. G. 1945, Rev. Mod. Phys., 17, 120.
    \item  Einstein, A., and Straus, E. G. 1946, Rev. Mod. Phys., 18, 148.
    \item  Ellis, G. F. R. 1971, in Relativistic Cosmology, Proc. of
           the International School of Physics ``Enrico Fermi'', General 
           Relativity and Cosmology, ed. Sachs, R. K. (Academic Press), p. 104.
    \item  Ellis, G. F. R., and Jaklitsch, M. J. 1989, Ap. J., 346, 601.
    \item  Feder, J. 1988, Fractals (New York: Plenum Press).
    \item  Hellaby, C., and Lake, K. 1985, Ap. J., 290, 381.
    \item  Hellaby, C., and Lake, K. 1986, Ap. J., 300, 461.
    \item  Israel, W. 1966, Nuovo Cimento, 44B, 1.
    \item  Israel, W. 1967, Nuovo Cimento, 48B, 463.
    \item  Jones, B. J. T., et al. 1988, Ap. J. (Letters), 332, L1.
    \item  Kantowski, R. 1969, Ap. J., 155, 1023.
    \item  Kristian, J., and Sachs, R. K. 1966, Ap. J., 143, 379.
    \item  Lake, K. 1980, Ap. J., 240, 744.
    \item  de Lapparent, V., Geller, M. J., and Huchra, J. P. 1986,
           Ap. J. (Letters), 302, L1.
    \item  MacCallum, M. A. H. 1983, in Relativistic Cosmology for
           Astrophysicists, The Origin and Evolution of Galaxies, eds. Jones,
           B. J. T., and Jones, J. E. (Dordrecht: D. Reidel and Co.), p. 9;
           also in Origin and Evolution of Galaxies, ed. de Sabbata, V.
           (Singapore: World Scientific, 1982), p. 9.
    \item  Mandelbrot, B. B. 1983, The Fractal Geometry of Nature
           (New York: Freeman).
    \item  Mart\'{\i}nez, V. J., et al. 1990, Ap. J., 357, 50.
    \item  Peebles, P. J. E. 1974, Ap. Space Sci., 31, 403.
    \item  Pietronero, L. 1987, Physica, 144A, 257.
    \item  Ruffini, R., Song, D. J., and Stoeger, W. R. 1988, Nuovo Cimento,
	   102B, 159.
    \item  Saunders, W., et al. 1991, Nature, 349, 32.
    \item  Tolman, R. C. 1934, Proc. Nat. Acad. Sci. (Wash.), 20, 169.
    \item  de Vaucouleurs, G. 1970, Science, 167, 1203.
    \item  Weinberg, S. 1972, Gravitation and Cosmology (John Wiley and Sons).
    \item  Wertz, J. R. 1971, Ap. J., 164, 227.
    \item  Wesson, P. S. 1978a, Ap. Space Sci., 54, 489.
    \item  Wesson, P. S. 1978b, Astr. Ap., 68, 131.
    \item  Wesson, P. S. 1979, Ap. J., 228, 647.
  \end{description}
\end{document}